\documentclass[twocolumn,amssymb, reprint,nobibnotes, showpacs, superscriptaddress, aps, pra]{revtex4-1}
\usepackage{graphicx}
\usepackage[tbtags]{amsmath}
\usepackage{epsfig}
\usepackage{natbib}
\usepackage{upgreek}

\begin{document}
\title{Suppressing phase decoherence of a single atom qubit with CPMG sequence}
\author{Shi Yu}
\affiliation{State Key Laboratory of Magnetic Resonance and Atomic and Molecular Physics, Wuhan Institute of Physics and Mathematics, Chinese Academy of Sciences - Wuhan National Laboratory for Optoelectronics, Wuhan 430071, China}
\affiliation{Center for Cold Atom Physics, Chinese Academy of Sciences, Wuhan 430071, China}
\affiliation{Graduate University of the Chinese Academy of Sciences, Beijing 100049, China}
\author{Peng Xu}
\affiliation{State Key Laboratory of Magnetic Resonance and Atomic and Molecular Physics, Wuhan Institute of Physics and Mathematics, Chinese Academy of Sciences - Wuhan National Laboratory for Optoelectronics, Wuhan 430071, China}
\affiliation{Center for Cold Atom Physics, Chinese Academy of Sciences, Wuhan 430071, China}
\author{Xiaodong He}
\affiliation{State Key Laboratory of Magnetic Resonance and Atomic and Molecular Physics, Wuhan Institute of Physics and Mathematics, Chinese Academy of Sciences - Wuhan National Laboratory for Optoelectronics, Wuhan 430071, China}
\affiliation{Center for Cold Atom Physics, Chinese Academy of Sciences, Wuhan 430071, China}
\author{Min Liu}
\affiliation{State Key Laboratory of Magnetic Resonance and Atomic and Molecular Physics, Wuhan Institute of Physics and Mathematics, Chinese Academy of Sciences - Wuhan National Laboratory for Optoelectronics, Wuhan 430071, China}
\affiliation{Center for Cold Atom Physics, Chinese Academy of Sciences, Wuhan 430071, China}\author{Jin Wang}
\affiliation{State Key Laboratory of Magnetic Resonance and Atomic and Molecular Physics, Wuhan Institute of Physics and Mathematics, Chinese Academy of Sciences - Wuhan National Laboratory for Optoelectronics, Wuhan 430071, China}
\affiliation{Center for Cold Atom Physics, Chinese Academy of Sciences, Wuhan 430071, China}
\author{Mingsheng Zhan}
\email{mszhan@wipm.ac.cn}
\affiliation{State Key Laboratory of Magnetic Resonance and Atomic and Molecular Physics, Wuhan Institute of Physics and Mathematics, Chinese Academy of Sciences - Wuhan National Laboratory for Optoelectronics, Wuhan 430071, China}
\affiliation{Center for Cold Atom Physics, Chinese Academy of Sciences, Wuhan 430071, China}
\begin{abstract}
We experimentally demonstrate the strong suppression of dephasing of a qubit stored in a single \mbox{\textsuperscript{87}Rb} atom in an optical dipole trap by using \mbox{Carr-Purcell-Meiboom-Gill} \mbox{(CPMG)} pulse sequences. Regarded as a repetition of spin echo, \mbox{CPMG} sequence is an optimal choice for suppressing both inhomogeneous and homogeneous phase decoherence mechanisms. In the trap with \mbox{830 nm} wavelength and \mbox{0.7 mK} potential depth, the spin relaxation time of single atoms is showed to be \mbox{830.8 ms}. We obtain the reversible inhomogeneous dephasing time of $T_{2}^{\ast}=1.4\text{ ms}$. The homogeneous dephasing time is $T_{2}^{\prime}=102.7\text{ ms}$ in the spin echo process, by employing \mbox{CPMG} sequence with pulse number $n = 6$ the homogeneous dephasing is further suppressed by a factor of 3, and its corresponding coherence time is extended to $T_{2}^{\prime}=304.5\text{ ms}$.
\end{abstract}
\maketitle

\section{Introduction}
Decoherence suppression in single neutral atoms is of great importance for quantum information processing \cite{Nielsen2000,Ladd2010} and precision measurements \cite {Snadden1998,Peters1999,Fixler2007}. Typically with trapped alkali\nobreakdash-metal atoms, the hyperfine levels of ground state can be considered as qubits, which are easily manipulated via microwave or Raman transitions. Multiqubit operations are performed based on dipole\nobreakdash-dipole interactions between highly excited Rydberg atoms \cite {Saffman2010}, \mbox{cavity-mediated} photon exchange \cite{You2003} and controlled ground-state collisions \cite{Jaksch1999}. These quantum systems have been employed to demonstrate a two-qubit controlled-NOT gate \cite{Isenhower2010} and the entanglement of two individual atoms \cite{Wilk2010}. Decoherence represents a loss of information and can be a serious limitation. Therefore coherence time need be considerably longer than initialization, multiqubit interaction and measurement times. Especially for scaling to plenty of atoms in the optical lattice, long coherence time is crucial to the storage and the manipulation of quantum information. For precision metrology, an interferometer consisting of only one atom is suitable to investigate the localized forces near surface \cite{Steffen2012,Parazzoli2012}. However, in the interferometer the effect of increasing sensitivity competes with the decreased ratio of signal to noise. Suppressing decoherence leads to prolongation of the measurement time while keeping large signal, and thus improves the precision.

Quantum decoherence is due to coupling between a quantum system and its surrounding environment, including the longitudinal spin relaxation ($T_{1}$) and the transverse phase relaxation ($T_{2}$). In a far-off-resonance trap \mbox{(FORT)}, the spin relaxation of atoms is due to the spontaneous Raman scattering of photons from the trap laser \cite{Cline1994}. A spin relaxation time of several seconds is achieved, owing to the large detuning. Therefore, a key issue is the improvement in phase coherence. As described in Ref. \cite{Kuhr2005}, it is important to distinguish between inhomogeneous and homogeneous processes. Inhomogeneous dephasing ($T_{2}^{\ast}$) originates from the energy distribution of trapped atoms which results in a distribution of light shift. In comparison with reversible inhomogeneous process, the homogeneous dephasing ($T_{2}^{\prime}$) affects each atom in the same way and cannot be reversed. Common mechanisms are the dipole trap laser intensity fluctuations, beam pointing instability, magnetic field fluctuations and heating of atoms. Multipulse sequence techniques have been investigated to refocus the phase diffusion and decouple the qubit from the environment. One of these, the \mbox{Carr-Purcell-Meiboom-Gill} \mbox{(CPMG)} pulse sequence is derived from spin echo and widely used in the field of nuclear magnetic resonance \cite{Slichter1990}. Recently, this technique has been applied to improve phase coherence of qubits in other systems, including semiconductor quantum dots \cite{Bluhm2011}, solid state superconductor \cite{Bylander2011}, nitrogen-vacancy centers in diamond \cite{Ryan2010}, atomic and ionic ensembles \cite{Andersen2004,Sagi2010,Biercuk2009}, and single ions \cite{Szwer2011}.

In this paper we investigate coherence of single \mbox{\textsuperscript{87}Rb} atoms in an optical dipole trap and strongly suppress dephasing utilizing the \mbox{CPMG} pulse sequence. For rubidium ensemble in the dipole trap, \mbox{CPMG} sequence with 10 $\pi$\nobreakdash-pulses was reported in Ref. \cite{Andersen2004}. The coherence time reached $T_{2}^{\prime}=65\text{ ms}$, a 2.5-fold improvement as compared to the value in the spin echo process. Later in a dense ensemble, utilizing \mbox{CPMG} sequence with more than 200 $\pi$\nobreakdash-pulses, the observed coherence time was $T_{2}^{\prime}=3\text{ s}$ \cite{Sagi2010}. For single neutral atoms, the spin echo technique was applied to reverse the inhomogeneous dephasing \cite{Kuhr2005,Kuhr2003,Jones2007,Beugnon2007}. Grangier's group obtained the dephasing times of $T_{2}^{\ast}=0.37\text{ ms}$ and $T_{2}^{\prime}=13\text{ ms}$ for the trap depth $U_{0}=1.2\text{ mK}$ \cite{Jones2007}. For $U_{0}=1.0\text{ mK}$, the dephasing time of $T_{2}^{\ast}=0.87\text{ ms}$ was reported by Saffman's group \cite{Yavuz2006}. However, CPMG sequence was not studied for single neutral atoms until the theoretical investigation of multipulse sequences in Ref. \cite{Chuu2009}. The authors examined the performance of variety of pulse sequences and found that \mbox{CPMG} pulse sequence was an optimal choice for suppressing both inhomogeneous and homogeneous dephasing mechanisms. To date, experimental demonstration of its effect remains unexplored. Here we extend CPMG sequence to single neutral atoms and successfully observe the strong suppression effect. For our trap $U_{0}=0.7\text{ mK}$, the spin relaxation time is $T_{1}=830.8\text{ ms}$. We obtain the inhomogeneous dephasing time of $T_{2}^{\ast}=1.4\text{ ms}$. Utilizing CPMG sequence with pulse number $n=6$, the homogeneous dephasing time is extended to $T_{2}^{\prime}=304.5\text{ ms}$. It is prolonged by a factor of 3 in comparison with the result of spin echo.

Our experimental setup and results are described in Sec. \ref{sec:experiment}. We measure the coherence time of single atoms and apply pulse sequences to suppress dephasing. In Sec. \ref{sec:theory} we theoretically analyze the suppression effect of CPMG sequence and fit experimental data according to the model. Finally, in Sec. \ref{sec:conclusion} we summarize the major findings and discuss the overall performance of quantum coherence in our system.

\section{\label{sec:experiment}Experimental tools and results}
\subsection{Experimental setup}

\begin{figure}[htbp]
\centering
\includegraphics [width=8.6cm]{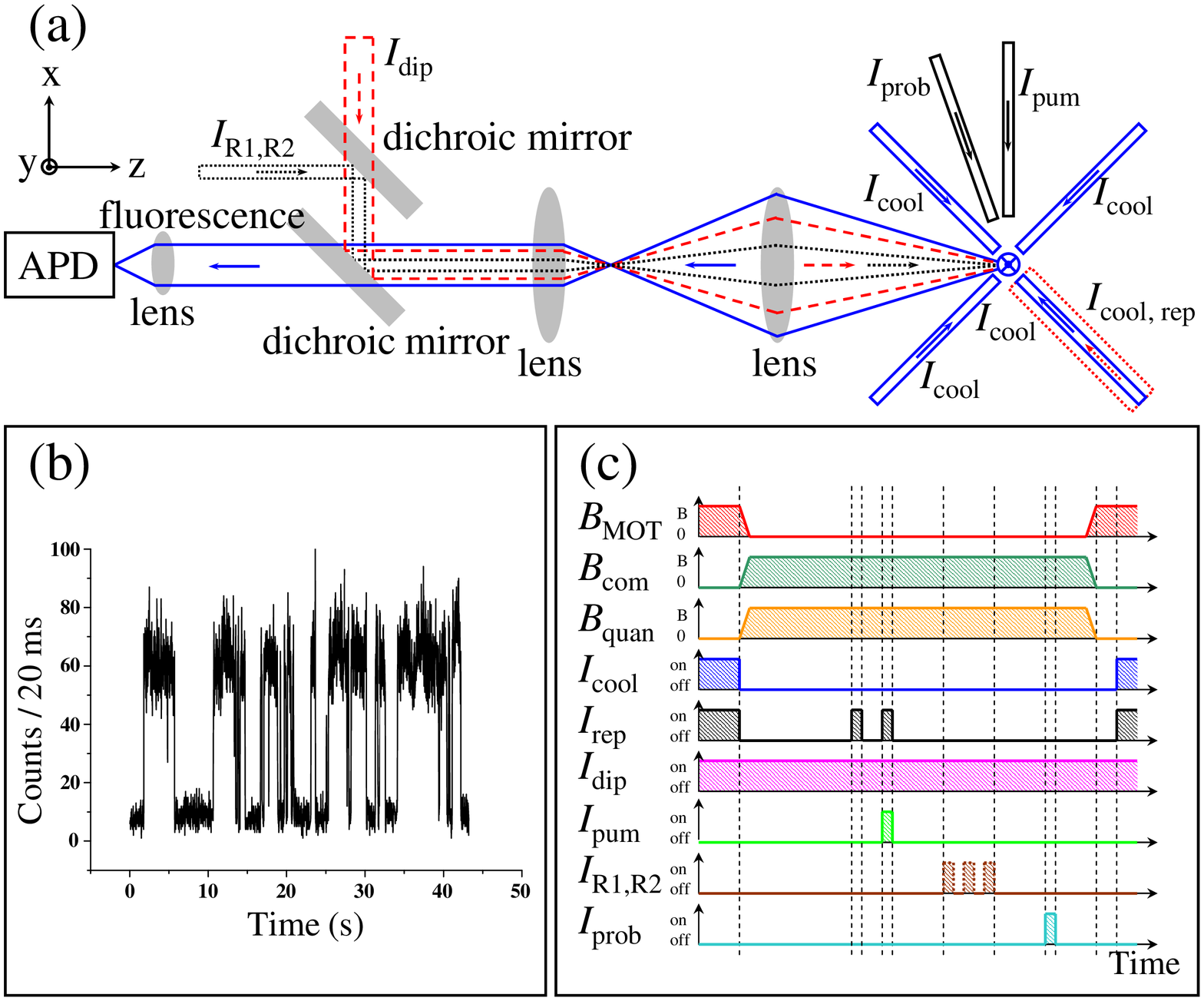}
\caption{(color online) (a) Schematic of experimental setup. The quantization axis is defined by a magnetic field along the \mbox{z-axis}. The \mbox{\textsuperscript{87}Rb} MOT is formed by six cooling lasers ($I_{\text{cool}}$) and an overlap repumping laser ($I_{\text{rep}}$). A laser beam with \mbox{830 nm} wavelength ($I_{\text{dip}}$) is tightly focused to a waist of \mbox{2.1 $\upmu\text{m}$}. The resulting optical potential depth is \mbox{0.7 mK} for a laser power of \mbox{7 mW}. The fluorescence of single atoms is detected with an avalanche photodiode. The initial state is prepared using an optical pumping laser beam ($I_{\text{pum}}$) in combination with the repumping beam. The Raman laser beams ($I_{\text{R1}},I_{\text{R2}}$) at \mbox{795 nm} are coupled into the same polarization maintaining optical fiber and focused onto single atoms. For the \mbox{state-selective} detection, a probe laser beam ($I_{\text{prob}}$) is applied. (b) Measured fluorescence signals of single atoms within \mbox{20 ms} time bins. Two steps correspond to either zero or one atom in the trap. (c) Experimental time sequence for dephasing investigation. We depict the MOT magnetic field, compensatory and quantization magnetic fields as $B_{\text{MOT}}$, $B_{\text{com}}$ and $B_{\text{quan}}$, respectively. See text for details.}
\label{fig:fig1}
\end{figure}

Our experimental setup is depicted in Fig. \ref{fig:fig1}(a) and has been described in detail elsewhere \cite{He2012}. Formed by six \mbox{counter-propagating} cooling lasers ($I_{\text{cool}}$) and an overlap repumping laser ($I_{\text{rep}}$), the magneto-optical trap (MOT) is loaded from background rubidium atoms in ultrahigh vacuum chamber. A laser beam propagating along z\nobreakdash-axis with \mbox{830 nm} wavelength ($I_{\text{dip}}$) is tightly focused in the center of the MOT by a commercial microscopic objective ($\text{N.A.}=0.38$). The resulting \mbox{far-off-resonance} trap (FORT) has a waist of \mbox{2.1 $\upmu\text{m}$}. Its potential depth is \mbox{0.7 mK} for a laser power of \mbox{7 mW}. By reducing the intensity and increasing the detuning of the MOT cooling beams, single atoms can be localized in the FORT based on \textquotedblleft collisional blockade \textquotedblright mechanism \cite{Schlosser2002}. This trap provides the storage time of about ten seconds in the absence of any near-resonant light. The lifetime is limited by collisions with background gas and the heating mechanisms due to intensity fluctuations and photon scattering of the trap laser. Measured by a \mbox{release-and-recapture} technique, the temperature of single atoms is about \mbox{40 $\upmu\text{K}$}. This value ensures that the mean kinetic energy of trapped atoms is much smaller than the potential depth. Therefore, the optical potential can be approximated by a harmonic oscillator, and the calculated axial and radial oscillation frequencies are $2\pi\times 3.5\text{ kHz}$ and $2\pi\times 39.2\text{ kHz}$. The laser induced fluorescence of single atoms is collected with the same objective, then coupled into a fiber and guided to an avalanche photodiode (APD) assembled in a single photon counting module. Shown in Fig. \ref{fig:fig1}(b), the fluorescence signals allow us to discriminate whether one atom is trapped or not.

\begin{figure}[htbp]
\centering
\includegraphics [width=8.6cm]{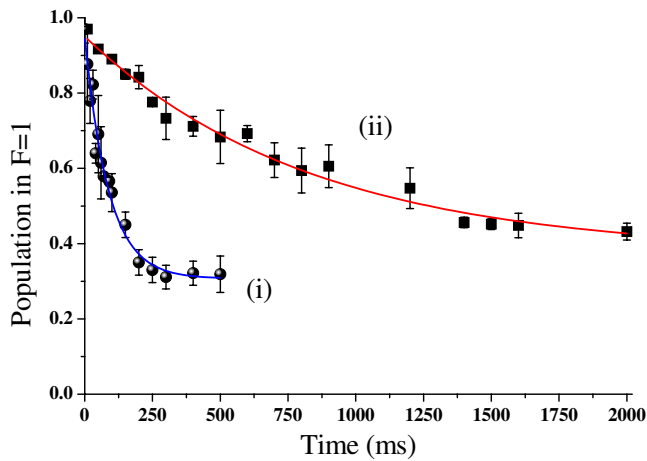}
\caption{(color online) Measurement of spin relaxation time. Single atoms is initialized in the state $|F=1\rangle$. After a variable trapping time, the push-out technique is applied to determine the atomic state. The fraction in $|F=1\rangle$ is recorded as a function of the trapping time. Each data is averaged over 200 single atoms. (i) Due to incoherent transitions induced by the background light of the dipole trap laser, the spin relaxation time is $T_{1}=87.1\pm9.4\text{ ms}$. (ii) With an interference filter to reduce the background, the obtained time is $T_{1}=830.8\pm114.7\text{ ms}$.}
\label{fig:fig2}
\end{figure}

\begin{figure}[htbp]
\centering
\includegraphics [width=8.6cm]{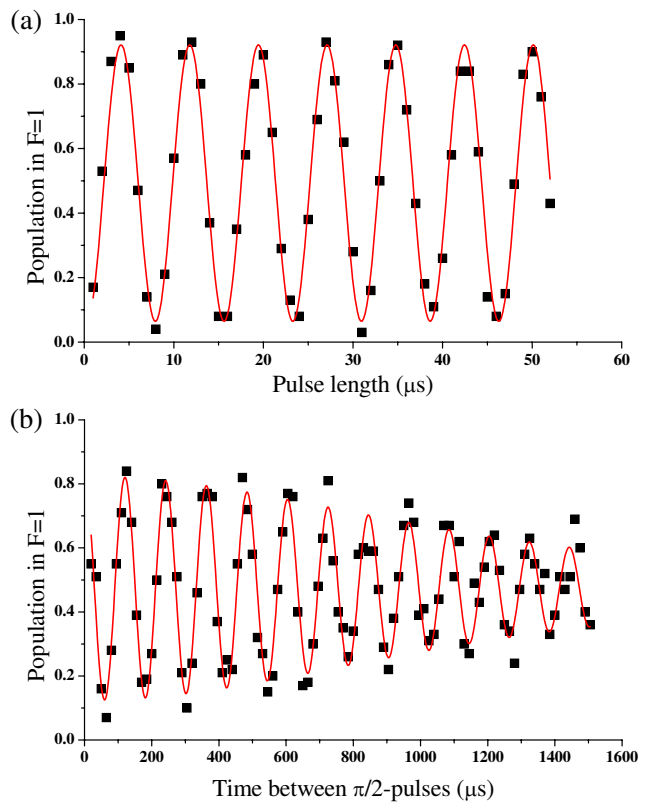}
\caption{(color online) (a) Rabi oscillations between $|F=1,m_{F}=0\rangle$ and $|F=2,m_{F}=0\rangle$ states. The fraction of single atoms in the state $F=1$ is measured as a function of Raman pulse length. We observe a sinusoidal variation with high contrast. The Rabi frequency is ¦¸$\Omega_{\text{R}}=2\pi\times 130\text{ kHz}$. This value corresponds to a $\pi/2$ rotation time of \mbox{1.92 $\upmu\text{s}$}. (b) Ramsey spectroscopy recorded with a fixed \mbox{two-photon} detuning from the \mbox{ground-state} hyperfine splitting. The detuning is ¦Ä$\delta=2\pi\times 8.6\text{ kHz}$. We fit the fringes according to the model presented in \cite{Kuhr2005}. The measured dephasing time is $T_{2}^{\ast}=1.4\text{ ms}$.}
\label{fig:fig3}
\end{figure}

\begin{figure}[htbp]
\centering
\includegraphics [width=8.6cm]{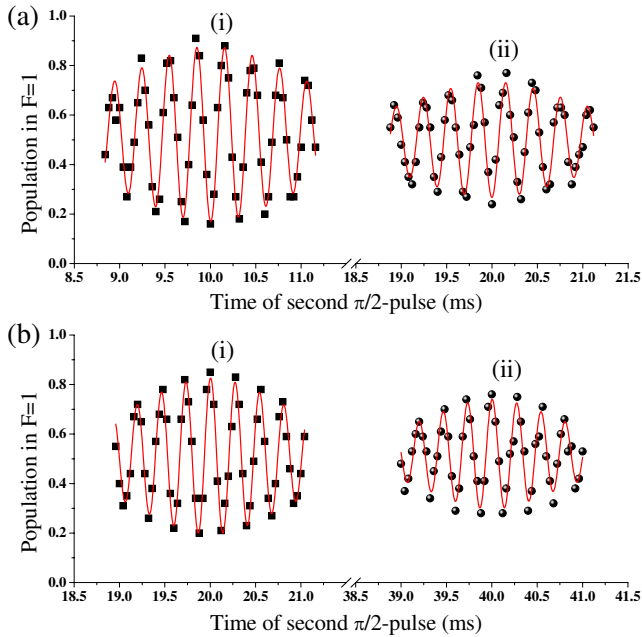}
\caption{(color online) Examples of the spin echo and CPMG signals. We plot fringes as a function of the time of second \mbox{$\pi/2$-pulse}. (a) For spin echo method, the time of \mbox{$\pi$-pulse} is fixed to be $\tau=5.0\text{ ms (i), }10.0\text{ ms (ii)}$. (b) For CPMG sequence with $n=2$, the first \mbox{$\pi$-pulse} is applied at time $\tau=5.0\text{ ms (i), }10.0\text{ ms (ii)}$, and thus the second \mbox{$\pi/2$-pulse} is at time $t=4\tau$. These graphs are fitted using the model in Ref. \cite{Kuhr2005} and in the text. The initial visibility decreases with increasing time $\tau$.}
\label{fig:fig4}
\end{figure}

\begin{figure}[htbp]
\centering
\includegraphics [width=8.6cm]{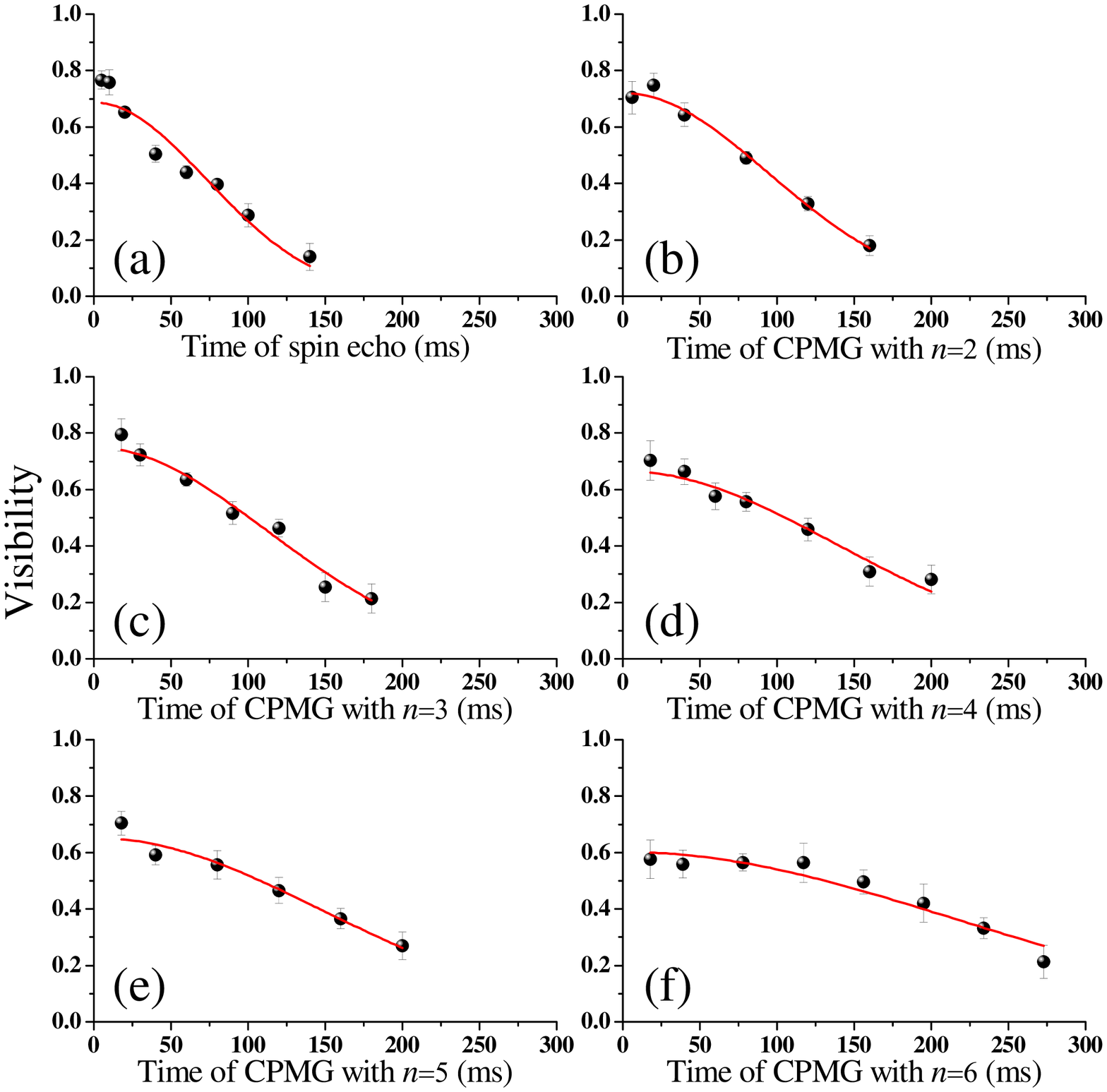}
\caption{(color online) Visibility of the spin echo and \mbox{CPMG} sequences as a function of the total time between two \mbox{$\pi/2$-pulses}. The values of spin echo (a) and CPMG sequences with pulse numbers $n=2\text{ (b), }3\text{ (c), }4\text{ (d), }5\text{ (e), and }6\text{ (f)}$ are extracted from the corresponding fringes similar to Fig.\ref{fig:fig4}. We fit all the graphs with a Gaussion depicted in Eq. (\ref{eq:eq1}) and (\ref{eq:eq16}).}
\label{fig:fig5}
\end{figure}

Single atoms can be prepared in $|F=2\rangle$ or $|F=1\rangle$ using the MOT laser beams. We define the logical states as $|1\rangle\equiv|F=2,m_{F}=0\rangle$ and $|0\rangle\equiv|F=1,m_{F}=0\rangle$. The quantization axis is generated by a bias magnetic field of $B=3\text{ G}$ along z\nobreakdash-axis. To compensate for the background magnetic field, three orthogonal pairs of coils are applied. The qubit is initialized in the state $|1\rangle$ by optically pumping. For this a $\pi$\nobreakdash-polarized laser beam ($I_{\text{pum}}$) propagates along x\nobreakdash-axis resonant with the $|F=2\rangle\rightarrow |F^{\prime}=2\rangle$ transition. Single qubit rotations between states $|0\rangle$ and $|1\rangle$ are performed by two-photon stimulated Raman transition. The Raman laser beams ($I_{\text{R1}},I_{\text{R2}}$) are formed by the $\pm1^{\text{st}}$ orders diffraction of a \mbox{3.4 GHz} \mbox{acoustic-optical} modulator (AOM). The zeroth order diffraction beam at \mbox{795 nm} is red detuned \mbox{60 GHz} from the $D_{1}$ transition. Both of the Raman beams are coupled into the same polarization maintaining optical fiber, and propagate along the quantization axis with circular polarization. They are focused in the chamber and overlapped with single atoms. The total power is \mbox{30 $\upmu\text{W}$}. For the \mbox{state-selective} detection, a circularly polarized probe laser beam ($I_{\text{prob}}$) resonant with the $|F=2\rangle\rightarrow|F^{\prime}=3\rangle$ transition is applied to push out single atoms in the state $|F=2\rangle$. However, atoms in the state $|F=1\rangle$ are not influenced by this laser and remain in the trap. The internal states are therefore mapped onto the presence of atoms.

\subsection{Spin relaxation measurement}
In order to measure the spin relaxation time, we switch off the MOT laser beams in proper sequence and prepare single atoms in the state $|F=1\rangle$. After a variable trapping time, the push-out laser beam is applied. We switch on the MOT beams again to determine whether the atom is still present. The fraction of single atoms in $|F=1\rangle$ is recorded as a function of the trapping time, as shown in Fig. 2. Each data is averaged over 200 single atoms.

In our experiment, the optical dipole trap at 830 nm wavelength is from an external cavity diode laser. It has a broad spectrum background light, including photons resonant with the rubidium $D_{1}$ and $D_{2}$ transitions. Due to incoherent transitions induced by the background, the obtained spin relaxation time is only $T_{1}=87.1\text{ ms}$. To eliminate this influence, we insert a narrow band interference filter (Semrock \mbox{FF01-830/2-25}) with central transmitting wavelength 830 nm. It has a full width at half maximum \mbox{(FWHM)} of 10.4 nm and a peak transmission of 95\%. The optical densities of 780 nm and 795 nm are close to 7. As a consquence, we almost completely remove the resonant photons from the optical dipole trap. The measurement spin relaxation time reaches $T_{1}=830.8\text{ ms}$. To compare with theory, we calculate the relaxation rate of the hyperfine-level populations using the Kramers-Heisenberg formula \cite{Cline1994}. Under our conditions, the theoretical spin relaxation time is about \mbox{909.0 ms}. The improved good agreement between theory and experiment is noted.

\subsection{Phase decoherence suppression}
The experimental sequence for dephasing investigation is shown in Fig. \ref{fig:fig1}(c). Once single atoms are trapped, we shut off the MOT laser beams and magnetic field (depicted as $B_{\text{MOT}}$), meanwhile turn on the compensatory and quantization magnetic field (depicted as $B_{\text{com}}$ and $B_{\text{quan}}$, respectively). After a \mbox{50 ms} interval to stabilize the magnetic fields, the qubit is initialized in state $|1\rangle$ using pumping and repumping laser beams. We then apply different Raman pulses to drive single qubit operation as desired, such as measuring dephasing time by Ramsey spectroscopy and suppressing dephasing via spin echo or CPMG pulses. At the end of the sequence, the MOT is switched on again, and the push\nobreakdash-out technique allows us to determine the projection of the atomic state on any superposition. For each data point plotted in Fig. \ref{fig:fig3} and Fig. \ref{fig:fig4}, we perform 100 cycles of this sequence.

Rabi oscillations on hyperfine levels of ground state are shown in Fig. \ref{fig:fig3}(a). We vary the Raman pulse length from \mbox{1 $\upmu\text{s}$} to \mbox{52 $\upmu\text{s}$} in steps of \mbox{1 $\upmu\text{s}$}. A sinusoidal variation with high contrast is observed. Fitted with a sinusoidal, we obtain a \mbox{two-photon} Rabi frequency of $\Omega_{\text{R}}=2\pi\times 130\text{ kHz}$, which corresponds to a $\pi/2$ rotation time of \mbox{1.92 $\upmu\text{s}$}.

The dephasing time is measured by performing Ramsey spectroscopy. We apply two $\pi/2$\nobreakdash-pulses with a fixed \mbox{two-photon} detuning $\delta$ separated by a variable time interval. During this time, atomic qubit evolves freely from a superposition of $|0\rangle$ and $|1\rangle$ states. As shown in Fig. \ref{fig:fig3}(b), we observe the Ramsey fringes with detuning $\delta=2\pi\times 8.6\text{ kHz}$. The decay of signal visibility as the free evolution time is due to the inhomogeneous dephaing. We fit the fringes using the model presented in \cite{Kuhr2005} and obtain the inhomogeneous dephasing time of $T_{2}^{\ast}=1.4\text{ ms}$.

This dephasing can be reversed by spin echo, which consists of an additional $\pi$\nobreakdash-pulse at half of the free evolution time in the Ramsey process. Examples of spin echo signals are shown in Fig. \ref{fig:fig4}(a) for different time between the first $\pi/2$\nobreakdash-pulse and $\pi$\nobreakdash-pulse, $\tau$. These fringes are plotted as a function of the time of second $\pi/2$\nobreakdash-pulse around $t=2\tau$. Obviously, the inhomogeneous dephasing has been reversed. We refer to \cite{Kuhr2005} and fit the spin echo signals. The visibility for different time of spin echo is recorded in Fig. \ref{fig:fig5}(a)and fitted with a Gaussion,
\begin{equation}
\label{eq:eq1}
V_{\text{spin echo}}(t)=C_{0}\exp\left[-\frac{1}{2}\left(\frac{t}{2}\right)^{2}\sigma_{\text{sig}}^{2}\right],
\end{equation}
where $\sigma_{\text{sig}}$ is the detuning fluctuation. The decrease of visibility results from the homogenous dephasing. Its corresponding phase decoherence time is $T_{2}^{\prime}=2\sqrt{2}/\sigma_{\text{sig}}=102.7\text{ ms}$ (shown in Table \ref{tab:tab1}). During the spin echo process, the homogenous dephasing due to low frequency noise is suppressed partially. Nevertheless, in the presence of higher frequency noise, multipulse sequence is a better choice for more effectively suppressing as well as compensating the phase error of $\pi$\nobreakdash-pulses \cite{Chuu2009}.

\begin{table*}
\caption{\label{tab:tab1}Fit parameters extracted from the signals of spin echo and CPMG sequence of Fig. \ref{fig:fig5} using Eq. (\ref{eq:eq1}) and (\ref{eq:eq16}). The parameter $n$ represents the number of \mbox{$\pi$-pulses} between two \mbox{$\pi/2$-pulses}. Spin echo is applied when $n=1$.}
\begin{ruledtabular}
\begin{tabular}{ccccccc}
&Fig. \ref{fig:fig5}(a)&Fig. \ref{fig:fig5}(b)&Fig. \ref{fig:fig5}(c)&Fig. \ref{fig:fig5}(d)&Fig. \ref{fig:fig5}(e)&Fig. \ref{fig:fig5}(f)\\
$n$&1&2&3&4&5&6\\\hline
$C_{0}$ (\%)&$68.7\pm2.9$&$72.1\pm1.7$&$74.9\pm2.4$&$66.6\pm2.0$&$65.2\pm2.3$&$60.2\pm1.7$\\
$\sigma_{\text{sig}}$ (Hz)&$27.6\pm2.0$&$42.4\pm1.3$&$53.5\pm2.8$&$57.4\pm3.2$&$67.5\pm4.3$&$55.7\pm3.1$\\
$T_{2}^{\prime}$ (ms)&$102.7\pm7.6$&$133.3\pm4.0$&$158.7\pm8.3$&$197.1\pm10.8$&$209.4\pm13.3$&$304.5\pm17.0$\\
\end{tabular}
\end{ruledtabular}
\end{table*}

Among different kinds of multipulse sequences, CPMG pulse, regarded as repetition of spin echo, is theoretically proved to be the most effective one for further prolonging the dephasing time \cite{Chuu2009}. It is composed of $n$ $\pi$\nobreakdash-pulses between two $\pi/2$\nobreakdash-pulses at time $t_{k}=(2k-1)\tau,k=1,2,\dotsc,n$, where $\tau$ is the time of the first $\pi$\nobreakdash-pulse. In our experiment we apply CPMG sequences with pulse numbers $n=2,3,4,5,\text{ and }6$. Similarly to the spin echo signals, CPMG fringes are obtained by scanning the time of the second $\pi/2$\nobreakdash-pulse. The corresponding signals are shown in Fig. \ref{fig:fig4}(b) and Fig. \ref{fig:fig5}(b)-(f).

\section{\label{sec:theory}Theoretical analysis of CPMG sequence}
We theoretically analyze the dephasing suppression effects of different CPMG sequences referring to \cite{Kuhr2005}. Resulting from the energy distribution of single atoms, a distribution of differential light shift $\delta_{\text{ls}}$ is given by
\begin{multline}
\label{eq:eq2}
p(\delta_{\text{ls}})=\frac{1}{2}\left(\frac{2\hbar\Delta_{\text{eff}}}{k_{B}T\omega_{\text{hfs}}}\right)^{3}
\left(\delta_{\text{ls}}-\delta_{0}\right)^{2}\\
\times\exp\left[-\frac{2\hbar\Delta_{\text{eff}}}{k_{B}T\omega_{\text{hfs}}}(\delta_{\text{ls}}-\delta_{0})\right],
\end{multline}
where $\delta_{0}$ is the maximum differential light shift, $\omega_{\text{hfs}}$ is the ground state hyperfine splitting, and $\Delta_{\text{eff}}$ is an effective detuning, including the effects of the $D_{1}$ and $D_{2}$ transitions. We define the matrices:
\begin{equation}
\label{eq:eq3}
{\bf\Phi}_{\pi/2}=
\begin{pmatrix}
1&0&0\\0&0&-1\\0&-1&0
\end{pmatrix},
\end{equation}
\begin{equation}
\label{eq:eq4}
{\bf\Phi}_{\pi}=
\begin{pmatrix}
1&0&0\\0&-1&0\\0&0&-1
\end{pmatrix},
\end{equation}
\begin{equation}
\label{eq:eq5}
{\bf\Psi}_{\text{free}}(\delta,t)=
\begin{pmatrix}
\cos(\delta t)&\sin(\delta t)&0\\-\sin(\delta t)&\cos(\delta t)&0\\0&0&1
\end{pmatrix}.
\end{equation}
They describe the action of $\pi/2$\nobreakdash-pulse and $\pi$\nobreakdash-pulse, the free precession with frequency $\delta$, respectively.

The initial state corresponds to the Bloch vector ${\bf U}_{0}\equiv(u,v,w)=(0,0,-1)$. After applying CPMG sequence, the Bloch vector evolves as
\begin{multline}
\label{eq:eq6}
{\bf U}_{\text{CPMG}}(t)={\bf\Phi}_{\pi/2}\times {\bf\Psi}_{\text{free}}\left(\delta,t-(2n-1)\tau\right)\times {\bf\Phi}_{\pi}\\
\shoveright{\times \underbrace{{\bf\Psi}_{\text{free}}(\delta,2\tau)\times {\bf\Phi}_{\pi}\times \dotsm \times {\bf\Psi}_{\text{free}}(\delta,2\tau)\times {\bf\Phi}_{\pi}}_{n-1 \text{ groups}}}\\
\times {\bf\Psi}_{\text{free}}(\delta,\tau)\times {\bf\Phi}_{\pi/2}\times {\bf U}_{0}.
\end{multline}
From this equation, we obtain the third component of the Bloch vector, $w$,
\begin{equation}
\label{eq:eq7}
w_{\text{CPMG}}(t)=(-1)^{n}\cos \left[\delta(t-2n\tau)\right].
\end{equation}
The shape of the CPMG fringe is calculated by integrating,
\begin{equation}
\label{eq:eq8}
\begin{split}
w_{\text{CPMG,inh}}(t)&=(-1)^{n}\int^{\infty}_{0}p(\delta_{ls})\\
&\phantom{1}\times\cos\left[\left(\delta_{\text{set}}-\delta_{\text{B}}-\delta_{\text{ls}}\right)(t-2n\tau)\right]d\delta_{\text{ls}}\\
&=(-1)^{n}\alpha\left(t-2n\tau,T^{\ast}_{2}\right)\\
&\phantom{1}\times\cos\left[\delta^{\prime}(t-2n\tau)+\kappa\left(t-2n\tau,T^{\ast}_{2}\right)\right],
\end{split}
\end{equation}
with
\begin{equation}
\label{eq:eq9}
\alpha\left(t-2n\tau,T^{\ast}_{2}\right)=\left[1+0.95(t-2n\tau)^{2}/T^{\ast}_{2}\right]^{-3/2},
\end{equation}
\begin{equation}
\label{eq:eq10}
\kappa\left(t-2n\tau,T^{\ast}_{2}\right)=-3\arctan\left[0.97(t-2n\tau)/T^{\ast}_{2}\right],
\end{equation}
where $\delta_{\text{set}}=\omega-\omega_{\text{hfs}}$ is the detuning, $\delta_{\text{B}}$ is the quadratic Zeeman shift,and $\delta^{\prime}=\delta_{\text{set}}-\delta_{\text{B}}-\delta_{\text{ls}}$ represents the sum of the detunings.
We fit the experimental data according to Eq. (\ref{eq:eq8}), as shown in Fig. \ref{fig:fig4}(b). All the measured visibility signals are plotted in Fig. \ref{fig:fig5}(b)-(f). In order to account for the homogenous dephasing of \mbox{CPMG} sequence, we introduce several average differences of detunings before and after the action of $\pi$\nobreakdash-pulses,
\begin{multline}
\label{eq:eq11}
\Delta\delta_{i}=\frac{1}{\tau}\left[\int^{2i\tau}_{(2i-1)\tau}\delta(t)dt-\int^{(2i-1)\tau}_{(2i-2)\tau}\delta(t)dt\right],\\
i=1,2,\dotsc,n.
\end{multline}
At time $t=2n\tau$, the CPMG sequence reads
\begin{multline}
\label{eq:eq12}
{\bf U}_{\text{CPMG}}(2n\tau)={\bf\Phi}_{\pi/2}\times{\bf\Psi}_{\text{free}}(\delta_{n}+\Delta\delta_{n},\tau)\\
\shoveright{\times{\bf\Phi}_{\pi}\times{\bf\Psi}_{\text{free}}(\delta_{n},\tau)
\times{\bf\Psi}_{\text{free}}(\delta_{n-1}+\Delta\delta_{n-1},\tau)}\\
\shoveright{\times{\bf\Phi}_{\pi}\times{\bf\Psi}_{\text{free}}(\delta_{n-1},\tau)
\times{\bf\Psi}_{\text{free}}(\delta_{n-2}+\Delta\delta_{n-2},\tau)}\\
\dotsm\\
\shoveright{\times{\bf\Phi}_{\pi}\times{\bf\Psi}_{\text{free}}(\delta_{2},\tau)
\times{\bf\Psi}_{\text{free}}(\delta_{2}+\Delta\delta_{2},\tau)}\\
\times{\bf\Phi}_{\pi}\times{\bf\Psi}_{\text{free}}(\delta_{1},\tau)\times{\bf\Phi}_{\pi/2}\times {\bf U}_{0}.
\end{multline}
We obtain
\begin{equation}
\label{eq:eq13}
w_{\text{CPMG}}(2n\tau)=(-1)^{n}\cos\left[\tau\sum^{n}_{i=1}(-1)^{n}\Delta\delta_{i}\right].
\end{equation}
All the detuning differences obey the Gaussian distribution with mean $\overline{\Delta\delta_{i}}=0$ and variance $\sigma_{i}$,
\begin{multline}
\label{eq:eq14}
p_{i}(\Delta\delta_{i})=\frac{1}{\sigma_{i}\sqrt{2\pi}}\exp\left[-\frac{(\Delta\delta_{i})^{2}}{2\sigma_{i}^{2}}\right],\\
i=1,2,\dotsc,n.
\end{multline}
Thus, the homogeneous component is depicted as
\begin{equation}
\label{eq:eq15}
\begin{split}
w_{\text{CPMG,hom}}(2n\tau)&=\idotsint_{-\infty}^{+\infty}w_{\text{CPMG}}(2n\tau)\\
&\phantom{1}\times\prod^{n}_{i=1}p_{i}(\Delta\delta_{i})
d\Delta\delta_{n}d\Delta\delta_{n-1}\dotsm d\Delta\delta_{\text{1}}\\
&=\exp\left(-\frac{1}{2}\tau^{2}\sum^{n}_{i=1}\sigma_{i}^{2}\right).
\end{split}
\end{equation}
We fit the visibility of CPMG sequence with a Gaussion similar to Eq. (\ref{eq:eq1}),
\begin{equation}
\label{eq:eq16}
V_{\text{CPMG}}(t)=C_{0}\exp\left[-\frac{1}{2}\left(\frac{t}{2n}\right)^{2}\sigma_{\text{sig}}^{2}\right].
\end{equation}
Here, we define $\sigma_{\text{sig}}$ by $\sigma_{\text{sig}}^{2}=\sum^{n}_{i=1}\sigma_{i}^{2}$.The homogenous dephasing time is defined as
\begin{equation}
\label{eq:eq17}
T_{2}^{\prime}=\frac{2\sqrt{2}n}{\sigma_{\text{sig}}}.
\end{equation}
The resulting fit parameters are listed in Table \ref{tab:tab1}. Note that the homogeneous dephasing time increases with number of CPMG pulses. For $n=2$, the dephasing time is \mbox{133.3 ms}. For $n=6$, the dephasing time is extended to \mbox{304.5 ms} and thus increased by a factor of 3 in comparison with the result of spin echo method.

\section{\label{sec:conclusion}Conclusions}
In conclusion, we have performed CPMG sequence with different pulse numbers in single \mbox{\textsuperscript{87}Rb} atoms and demonstrated the strong suppression of dephasing. We analyze the effect of CPMG sequence and experimentally confirm that it could efficiently suppress both inhomogeneous and homogeneous dephasing mechanisms.

For our trap $U_{0}=0.7\text{ mK}$, the homogeneous dephasing time reaches $T_{2}^{\prime}=304.5\text{ ms}$ via CPMG sequence with $n=6$. It is already prolonged by a factor of 3. Once we increase pulse number $n>6$, the initial visibility decreases significantly, owing to imperfections of the pulse generation and the state preparation. This influence limits our suppression effect, while it could be improved in the future. In combination with the improvement in stabilities of the magnetic field, the trap laser intensity and beam pointing, we will obtain the longer homogeneous dephasing time. Via spin echo and CPMG pulse sequence, the inhomogeneous dephasing is successfully reversed. However, its corresponding time remains $T_{2}^{\ast}=1.4\text{ ms}$. This results from unchanged atomic energy distribution. We can reduce inhomogeneous broadening utilizing the adiabatic lowering of the trap depth and Raman sideband cooling techniques. Under our experimental conditions, the obtained spin relaxation time is $T_{1}=830.8\text{ ms}$. The measurement time is consistent with the calculated value. For quantum information processing, fundamentally one needs $T_{2}\leq2T_{1}$ \cite{Ladd2010}. Thus, if longer dephasing time is expected, we have to further suppress the spin relaxation of atoms, for example, an optical dipole trap with larger detuning and smaller depth can be used.

Moreover, in comparison with the \mbox{red-detuned} trap, photon scattering and motional decoherence effects are significantly reduced in the \mbox{blue-detuned} trap \cite{Xu2010}. The coherence time $T_{2}^{\ast}=43\text{ ms}$ of single cesium atoms was observed in a 532 nm wavelength bottle beam trap \cite{Li2012}. The suppression effect of CPMG sequence for single atoms in the dark trap is anticipated. We believe this method will be useful in further experiments of quantum information processing and precision measurements with single neutral atoms.

\begin{acknowledgments}
This work was supported by the National Basic Research Program of China under Grant No.2012CB922101, the National Natural Science Foundation of China under Grant Nos.11104320 and 11104321, and funds from the Chinese Academy of Sciences.
\end{acknowledgments}

\end{document}